# Pediatric Bone Age Assessment using Deep Learning Models

Aravinda Raman[1], Sameena Pathan[1], Tanweer Ali[2]
[1] Dept. of Information and Communication Technology, Manipal Institute of Technology, MAHE, Manipal
[2] Dept. of Electronics and Communication Engineering, Manipal Institute of Technology, MAHE, Manipal
aravindaraman04@gmail.com, sameena.bp@manipal.edu, tanweer.ali@manipal.edu

*Abstract* — **Bone age assessment (BAA) is a standard method for determining the age difference between skeletal and chronological age. Manual processes are complicated and necessitate the expertise of experts. This is where deep learning comes into play. In this study, pre-trained models like VGG-16, InceptionV3, XceptionNet, and MobileNet are used to assess the bone age of the input data, and their mean average errors are compared and evaluated to see which model predicts the best.**

## 1. INTRODUCTION

Many challenges in various domains have experienced breakthrough technologies with new and novel solutions because of Machine Learning, Image Processing, and Statistical Learning improvements. Medical imaging has received much attention from the machine learning community, which has resulted in new ways of solving old problems. Predicting bone age from a sequence of x-ray images is an issue that our research is particularly interested in. The goal is to forecast the bone age within a year tolerance using a training set of x-rays of an individual's hands and related gender. Finding high-level descriptors that appropriately reflect bone age is one of the issues associated with this. Moreover, there is no denying that a person's gender impacts the outcome.

## 2. METHODOLOGY

*A. Dataset*

The dataset used in this study was first utilized in the Radiological Society of North America (RSNA) 2017 competition. Hand X-ray scans for people aged 1 to 288 months are included in this dataset. The training set includes 12611 unique hand scans identified with the owner's age and gender. Figure 1 shows a selection of photos from the training image collection. Figure 2 shows the age distribution (in months) of boys and girls; Figure 3 depicts the entire population's age distribution (in months). Figure 4 illustrates the count of men and women. As shown in Figure 3, only a few samples lie near the two borders, whereas most samples have an age of around 150 months. The gender distribution is also unbalanced, with 6833 males and 5778 females, as shown in Figure 4.

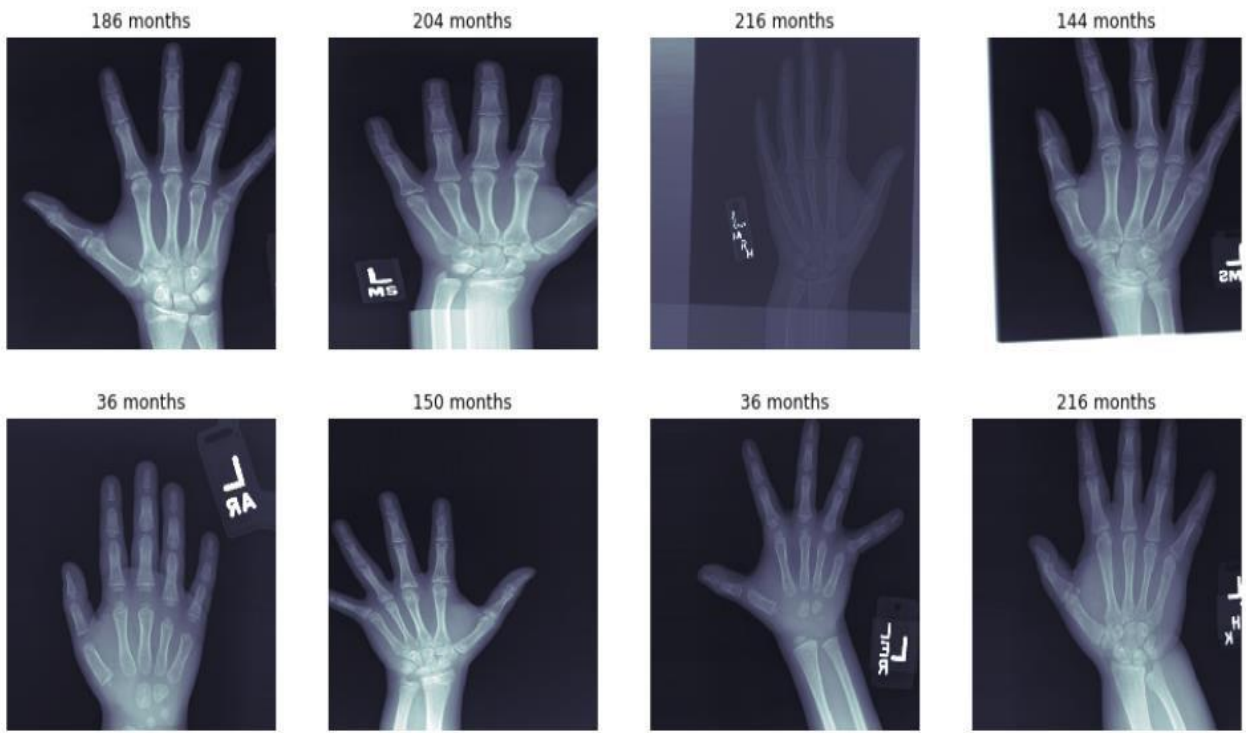

**Fig 1. RSNA dataset samples**

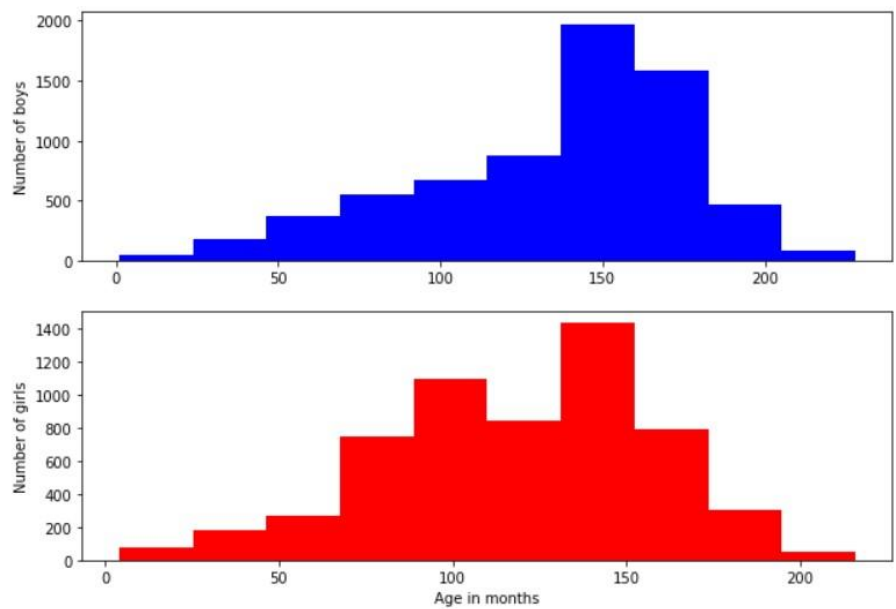

**Fig 2. Bone age distribution (boys and girls)**

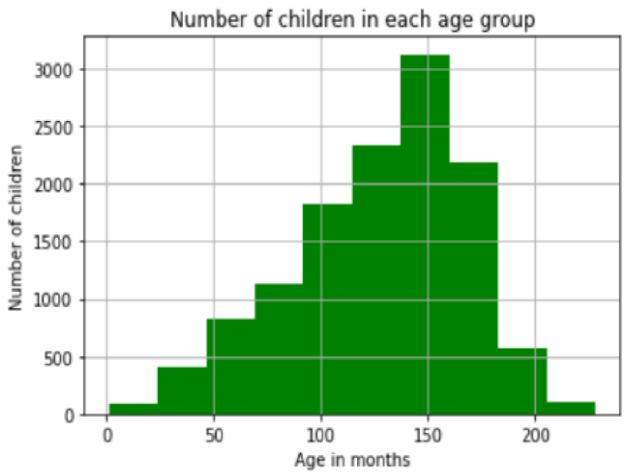 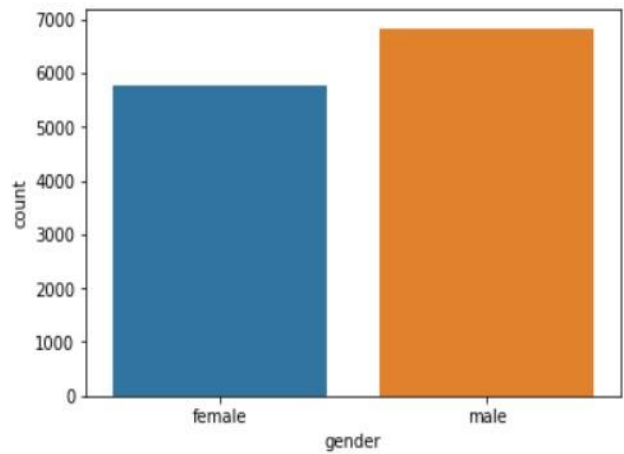

**Fig 3. Bone age distribution**  **Fig 4. Gender Distribution**

*B. Bone age assessment model:*

The data is split into 6000 training and 2000 validation images, i.e., a test size of 25% is used. The first step after obtaining a sample set of images is pre-processing. Here, the images are pre-processed using the appropriate pre-processing input, and the appropriate size of the image is used for each pretrained model. Image augmentation is explicitly used here because the dataset images predominantly contain left hands, and the training set images are randomly flipped horizontally to add complexity to the dataset. Shearing and zooming are used on the photos for the same reason.

Furthermore, the hand in the photograph may be crooked. Therefore, the deep learning model is given a small rotation range for the image. All these steps lead to an improvement in bone age prediction.

After pre-processing, the second step is to get the ground truth labels (age in months) for each training image and pass them to the CNN model for training.

The third step involves the model training with the pre-processed images with their respective ground truth labels. Transfer learning techniques are used here to train the models. VGG-16, InceptionV3, MobileNet, and XceptionNet are the models used here.

Transfer learning is usually accomplished in one of several methods:

1) Reusing the model architecture and pre-trained weights (in this case, 'ImageNet' weights are used) as an initialization point, retraining the complete network from start to finish. This has the advantage of producing a model tailored to a specific problem and takes advantage of the generality of a deep network's first layers.

2) Using the same model architecture but applying the pre-trained weights (ImageNet weights) to the first set of layers and freezing them during the training process, leaving just the last layers trained. This usually takes advantage of the fact that most of a deep network's initial levels may be regarded as a set of feature extraction layers, with the final layers doing the task at hand. As a result, this technique efficiently employs a learned feature extract while training an associated network to do a different task.

3) Another transfer learning method is to use pre-trained weights, freeze the first set of layers, and augment the final layers with a new architecture tailored to a given situation.

*C. Model Implementation*

Method 1: Transfer learning method 1 is used here, i.e., the model architecture is kept the same, and the entire model is trained starting from the first till the last layer, but the final SoftMax layer of all the pre-trained models is removed and is augmented with Batch Normalization, Global Average Pooling, and Dropout layers before the fully connected layer to prevent overfitting and reduce the mean average error. Finally, an output layer with a single neuron is used to get the regression results.

Method 2: Transfer learning method 2 is employed, i.e., the same model architecture is used along with the pre-trained weights for initialization. The first set of layers is frozen, leaving only the last layers to train. The final SoftMax layer is removed and augmented with Batch Normalization, Global Average Pooling, and Dropout layers before the fully connected layer. The last layer consists of a single neuron which gives the regression output. Both method 1 and method 2 have the same architecture. Only how the models are trained differs. 'Adam' optimizer is used for all the models, 'mean squared error' (MSE) as the loss function and 'mean average error' (MAE) as the metric. Each model is trained for 15 epochs with a patience of 10. Here patience refers to the number of epochs of no improvement in the Validation-MAE, after which the model's training ceases. Finally, the trained model is used to make predictions on the test dataset, which contains 200 images, and the bone age assessment result is obtained.

The proposed classification architectures are described as follows:

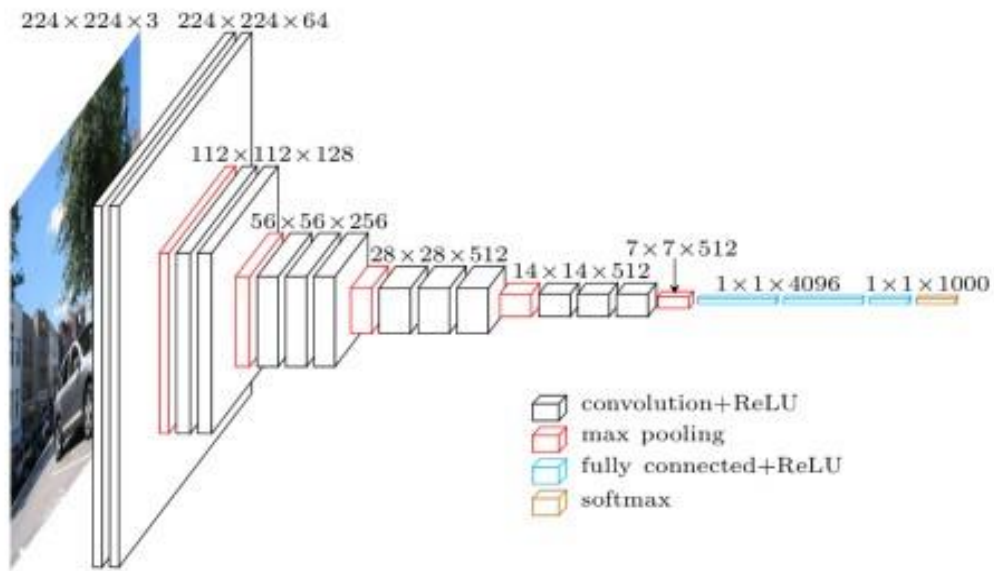

Fig 5. VGG-16 Architecture [1]

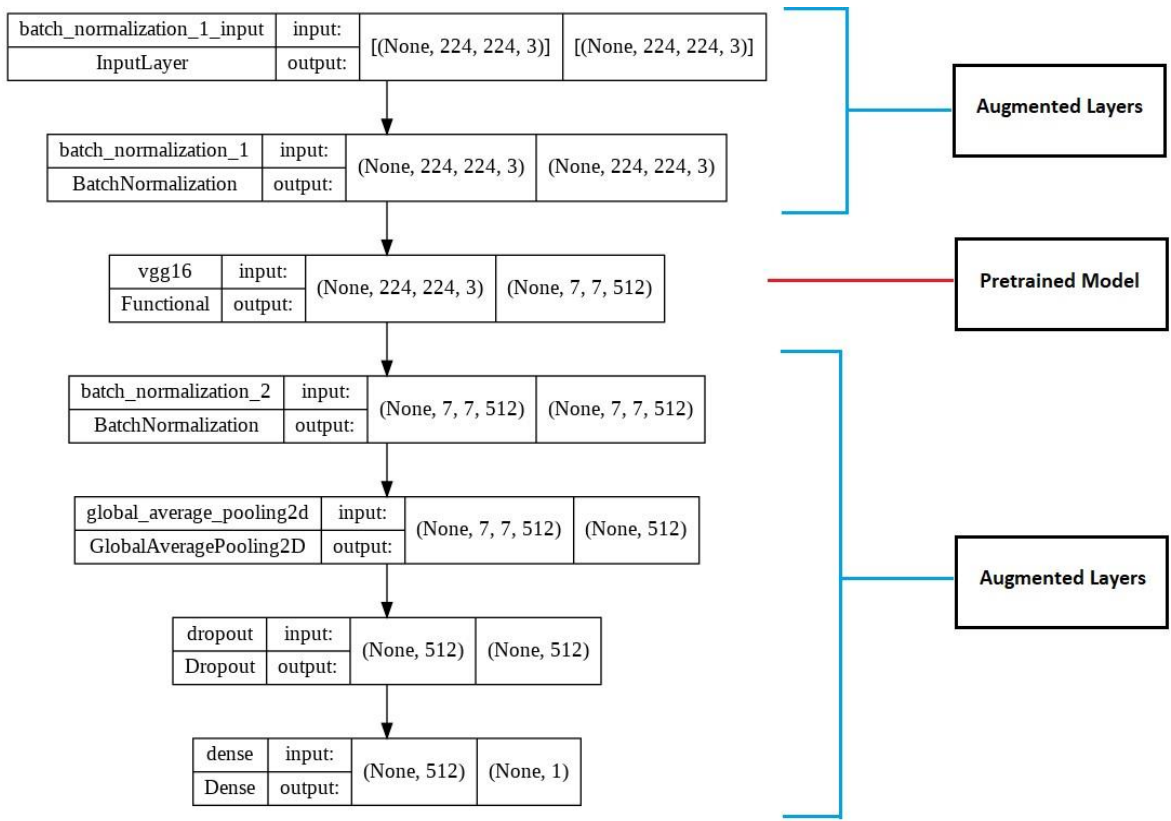

Fig 6. Model Architecture of VGG-16

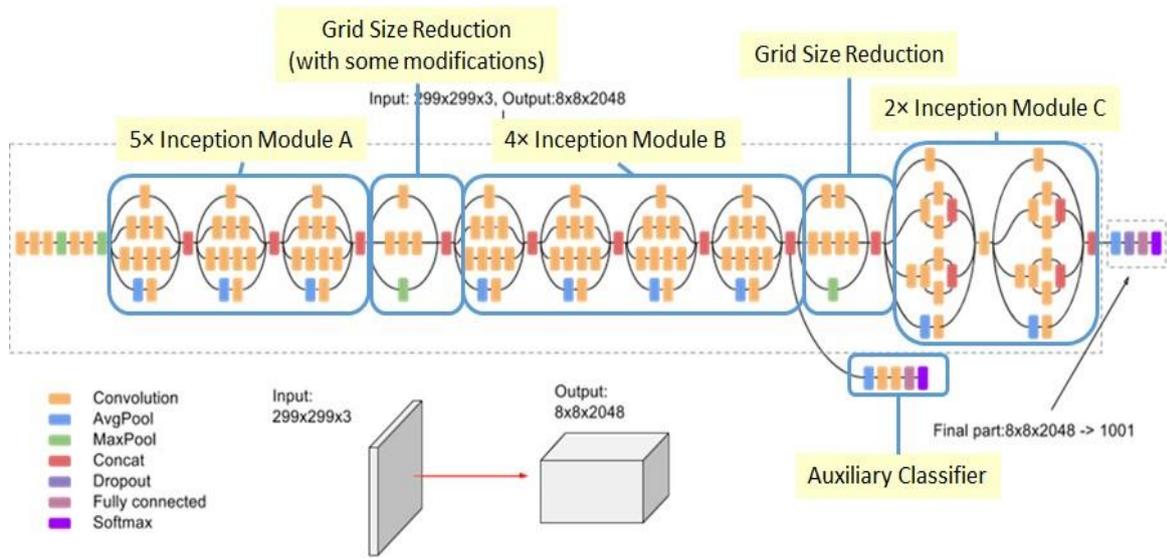

Fig 7. InceptionV3 Architecture [2].

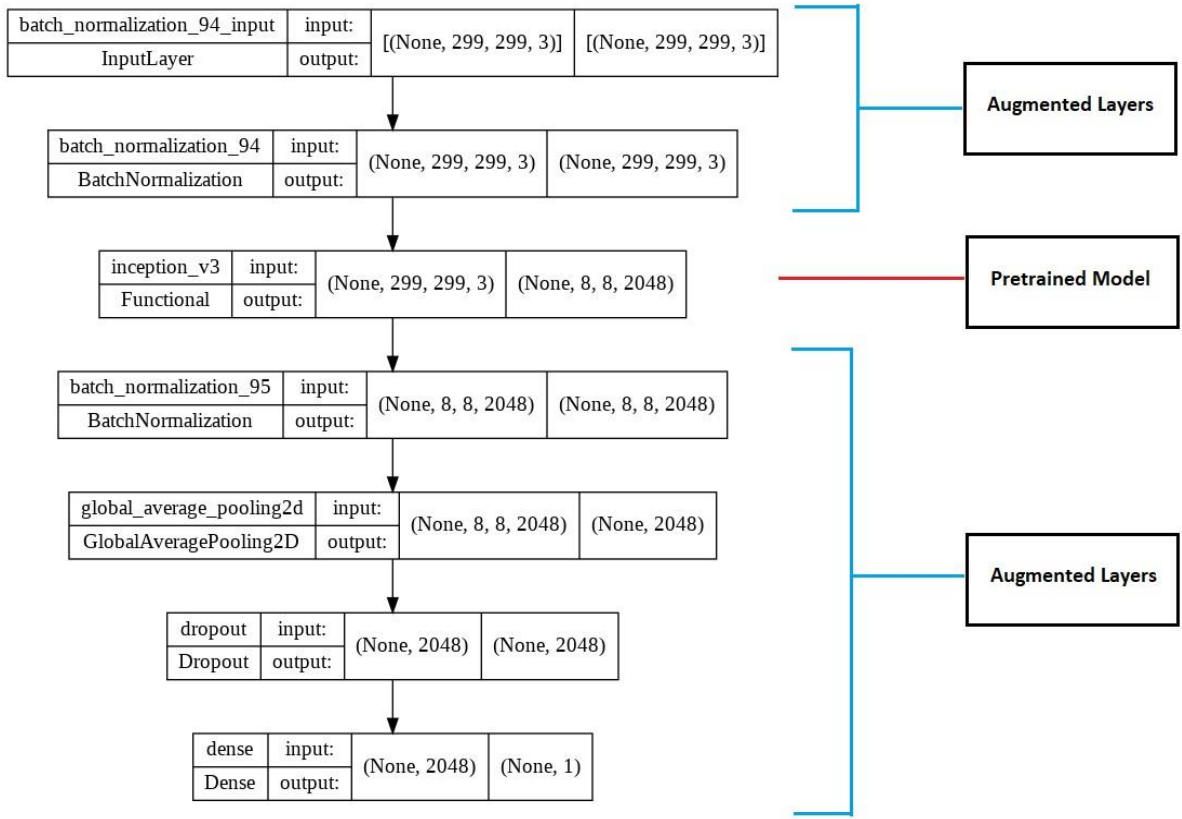

Fig 8. Model Architecture of InceptionV3

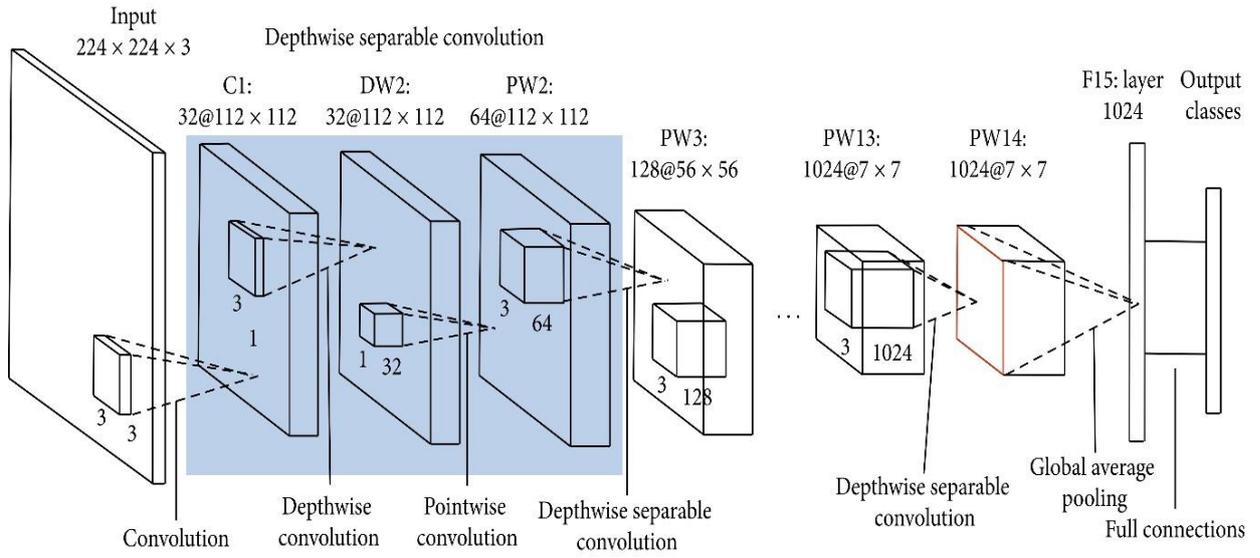

**Fig 9. MobileNet Architecture [3]**

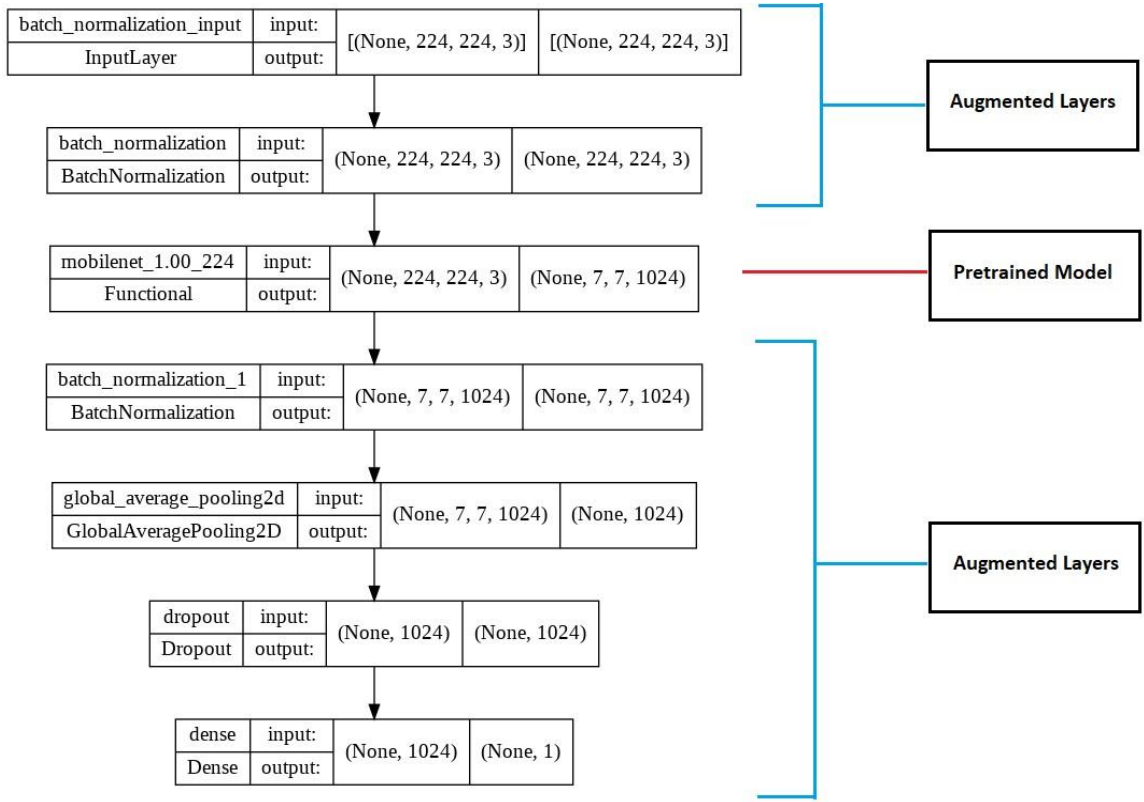

**Fig 10. Model Architecture of MobileNet**

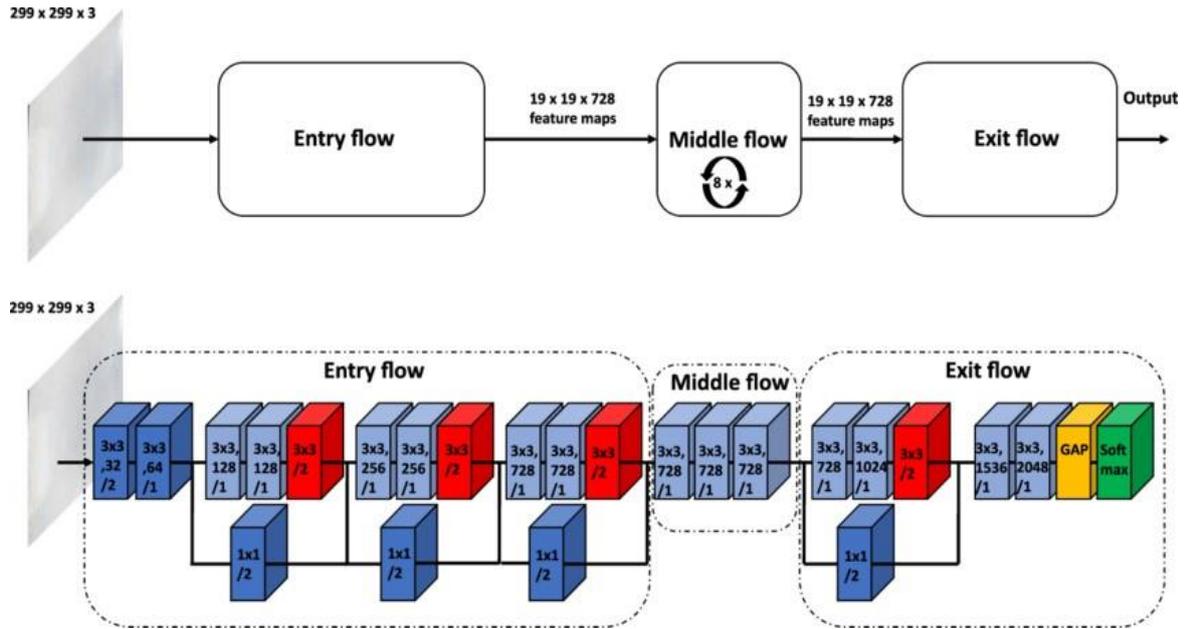

**Fig 11. XceptionNet Architecture [4].**

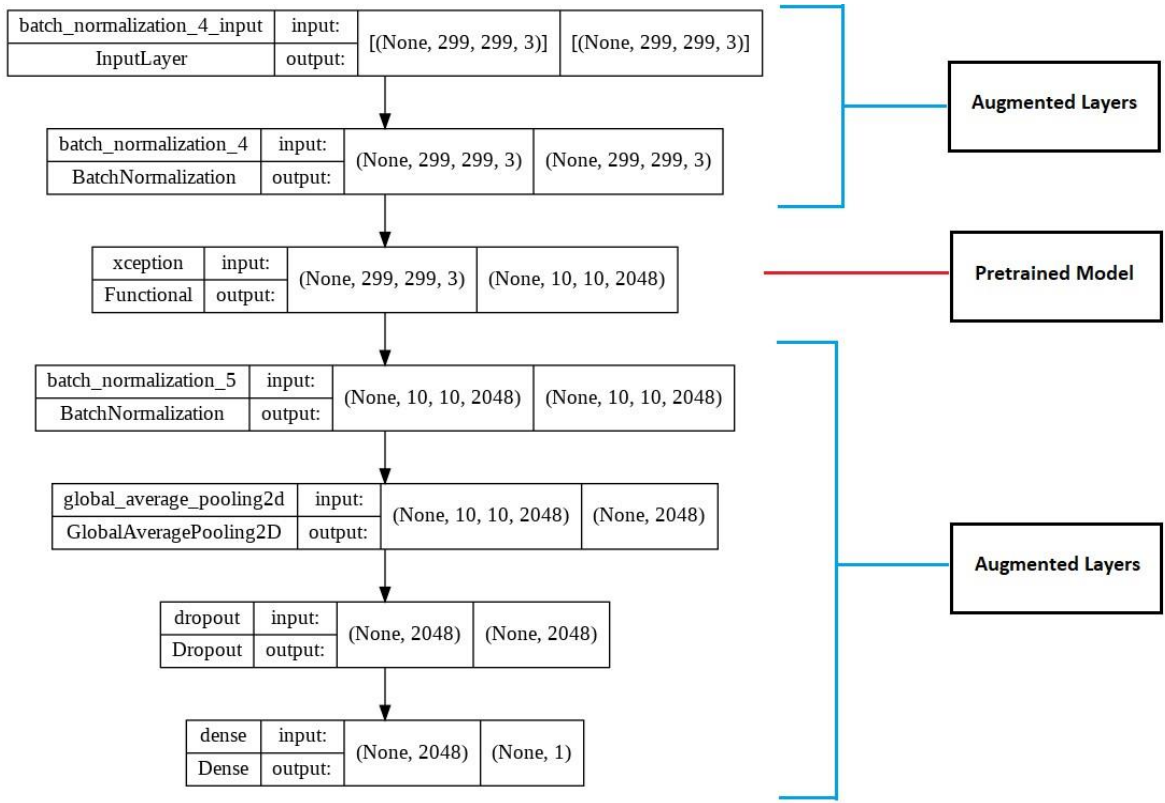

**Fig 12. Model Architecture of XceptionNet**

# 3. RESULTS

*A. VGG-16*

Figure 13 shows the mean average error variation for the train and test images, and Figure 14 shows the model's prediction versus the actual data.

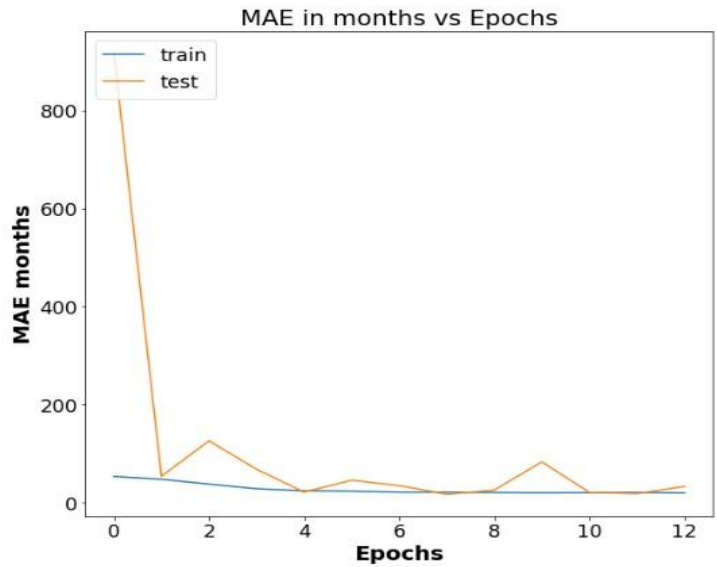

Fig 13. MAE months vs. Epochs

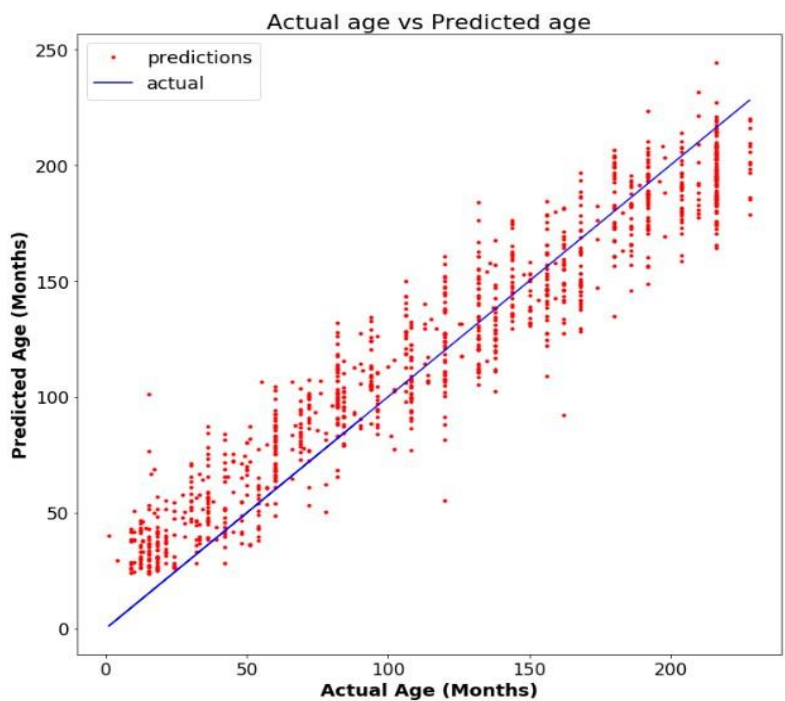

Fig 14. Actual age vs. Predicted age in months

A mean average error (MAE) of 14 months is obtained on test data.

**Results of model training *using method 2*:**

Figure 15 shows the mean average error variation for the train and test images, and Figure 16 shows the model's prediction versus the actual data.

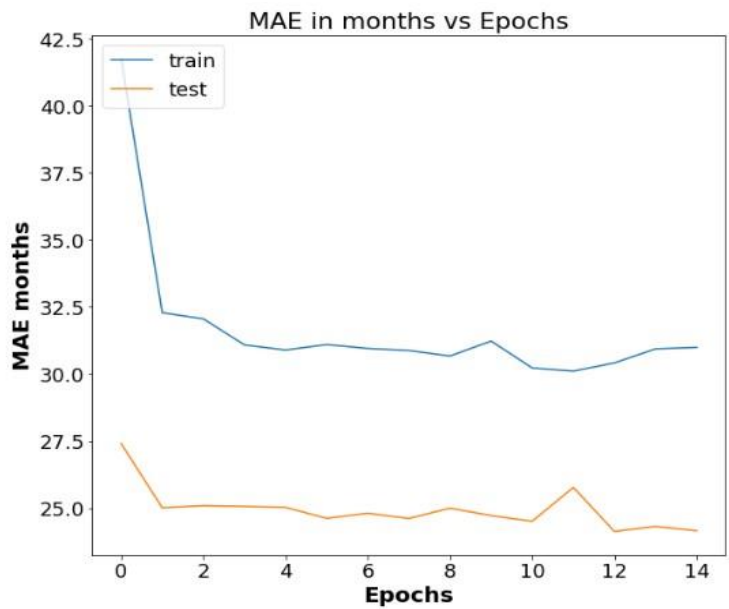

**Fig 15. MAE months vs. Epochs**

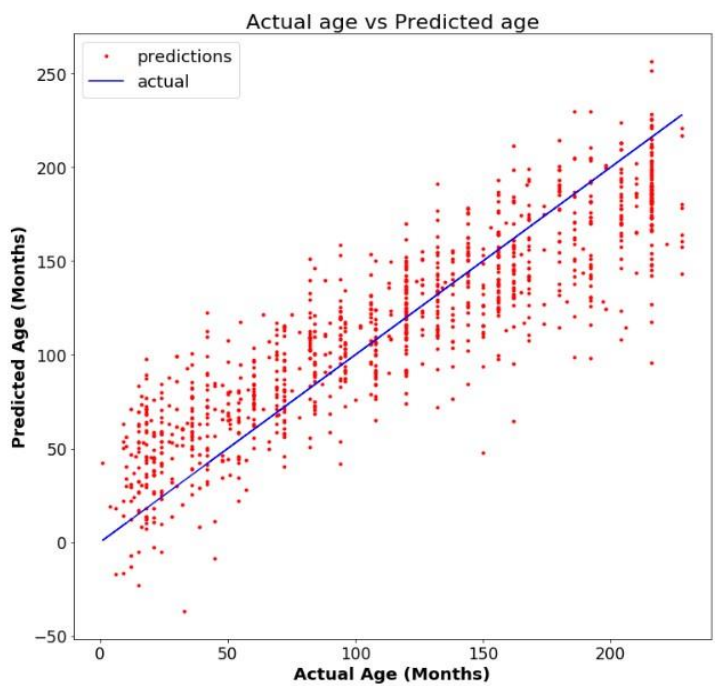

**Fig 16. Actual age vs. Predicted Age in months**

A mean average error (MAE) of 29.51 months is obtained on test data.

*B. Inception V3*

Figure 17 shows the mean average error variation for the train and test images, and Figure 18 shows the model's prediction versus the actual data.

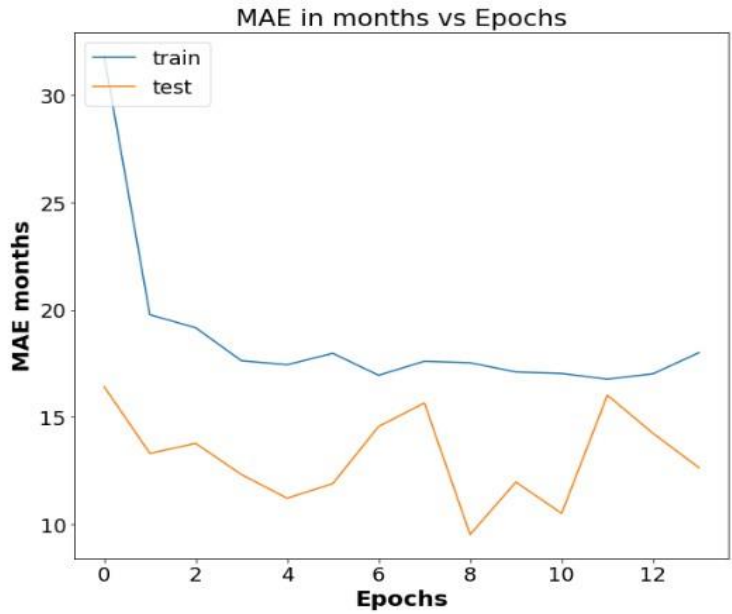

**Fig 17. MAE months vs. Epochs**

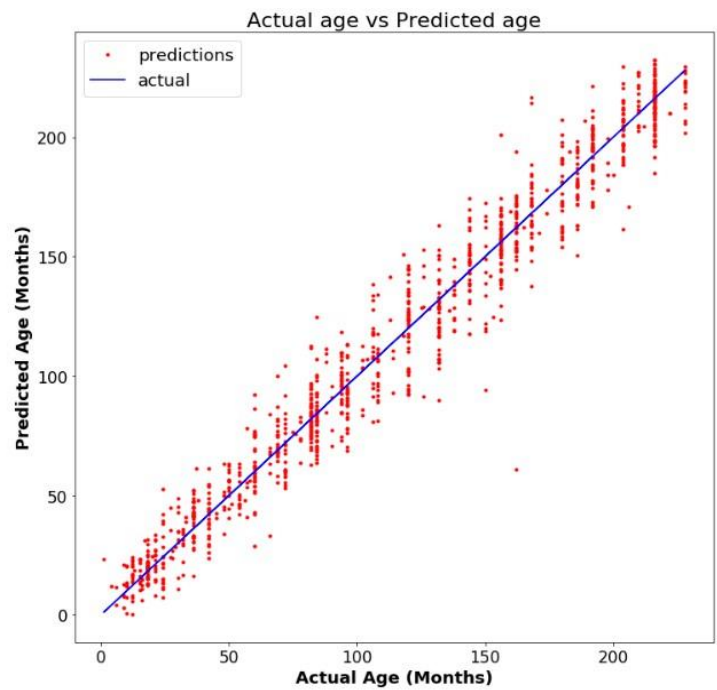

**Fig 18. Actual age vs. Predicted Age in months**

A mean average error (MAE) of 10.23 months is obtained on test data.

**Results of model training *using method 2*:**

Figure 19 shows the mean average error variation for the train and test images, and Figure 20 shows the model's prediction versus the actual data.

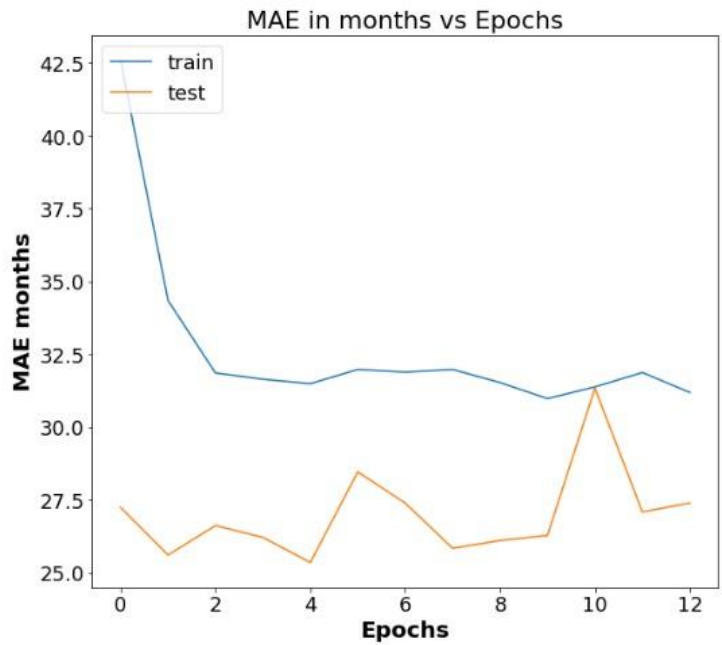

**Fig 19. MAE months vs. Epochs**

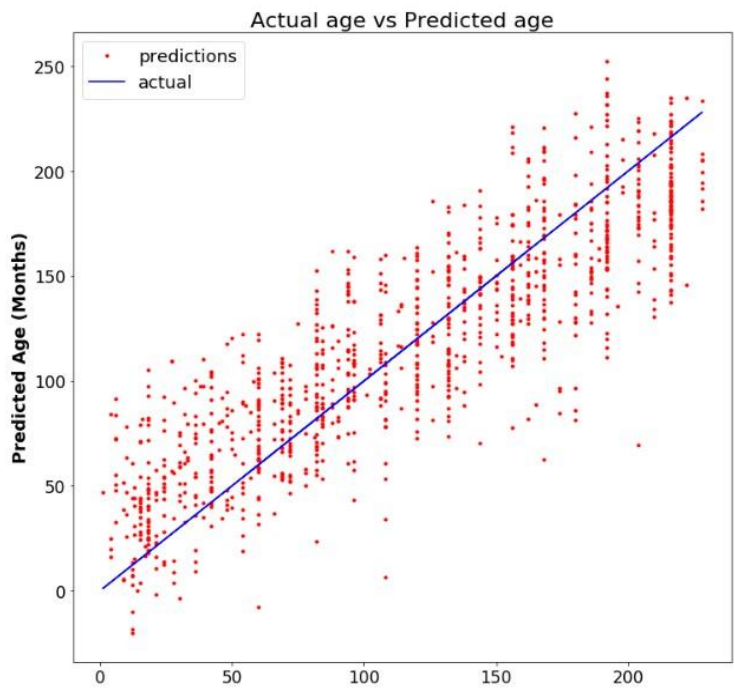

**Fig 20. Actual age vs. Predicted Age in months**

A mean average error (MAE) of 29.13 months is obtained on test data.

*C. MobileNet*

Figure 21 shows the mean average error variation for the train and test images, and Figure 22 shows the model's prediction versus the actual data.

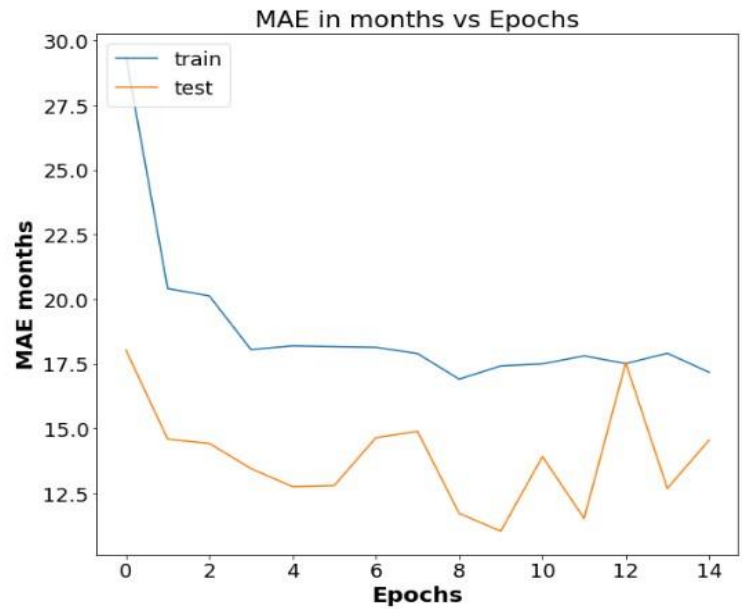

**Fig 21. MAE months vs. Epochs**

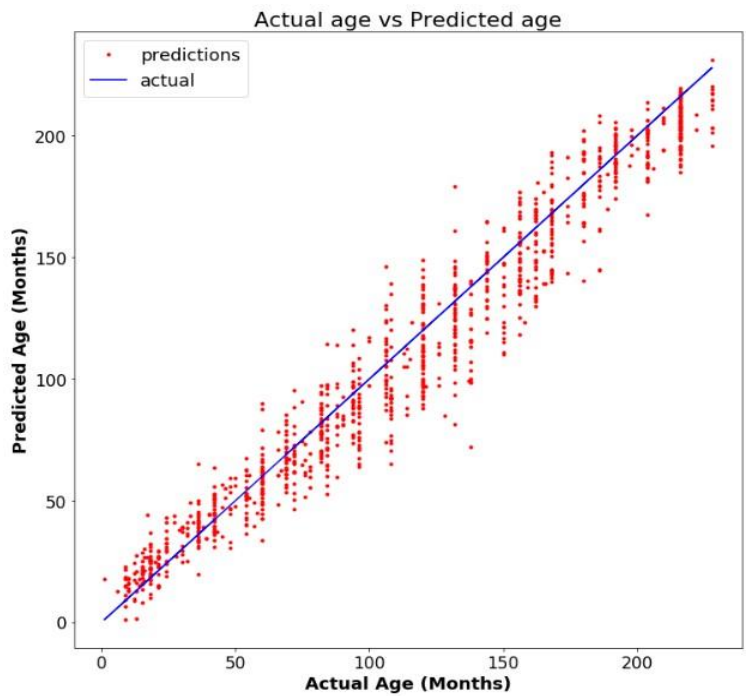

**Fig 22. Actual age vs. Predicted Age in months**

A mean average error (MAE) of 9.55 months is obtained on test data

**Results of model training *using method 2*:**

Figure 23 shows the mean average error variation for the train and test images, and Figure 24 shows the model's prediction versus the actual data.

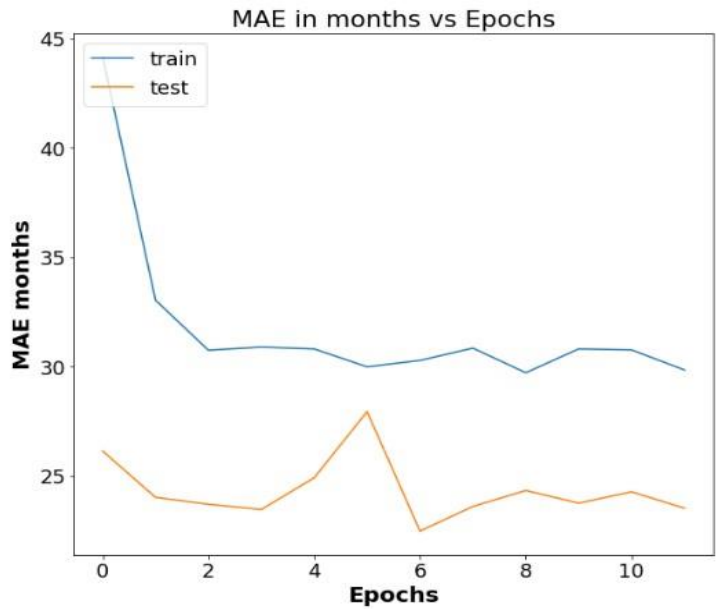

**Fig 23. MAE months vs. Epochs**

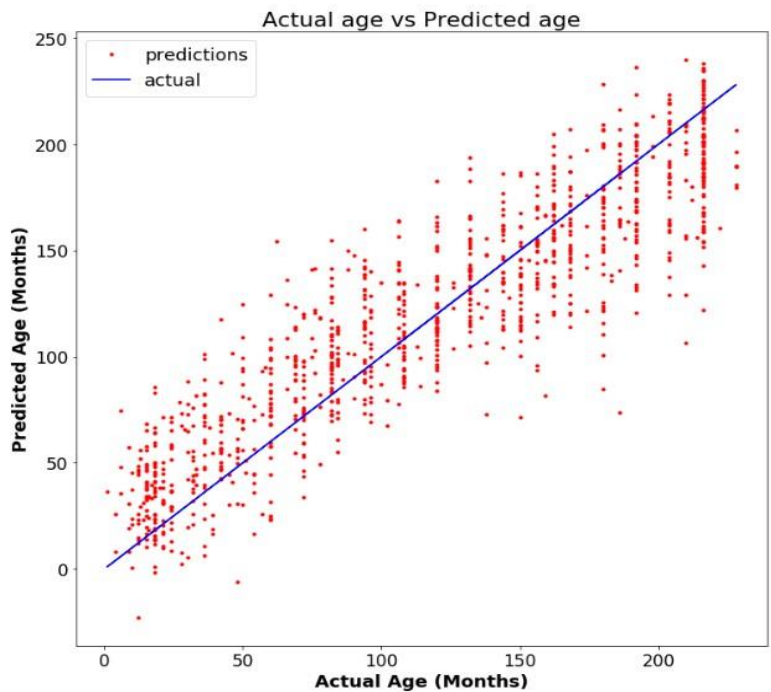

**Fig 24. Actual age vs. Predicted Age in months**

A mean average error (MAE) of 29.56 months is obtained on test data.

*D. XceptionNet*

Figure 25 shows the mean average error variation for the train and test images, and Figure 26 shows the model's prediction versus the actual data.

**Results of model training *using method 1*:**

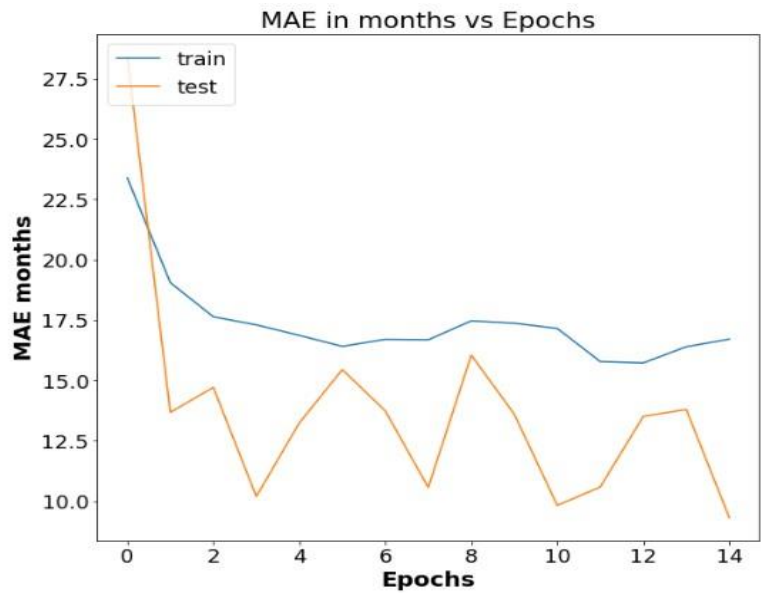

**Fig 25. MAE months vs. Epochs**

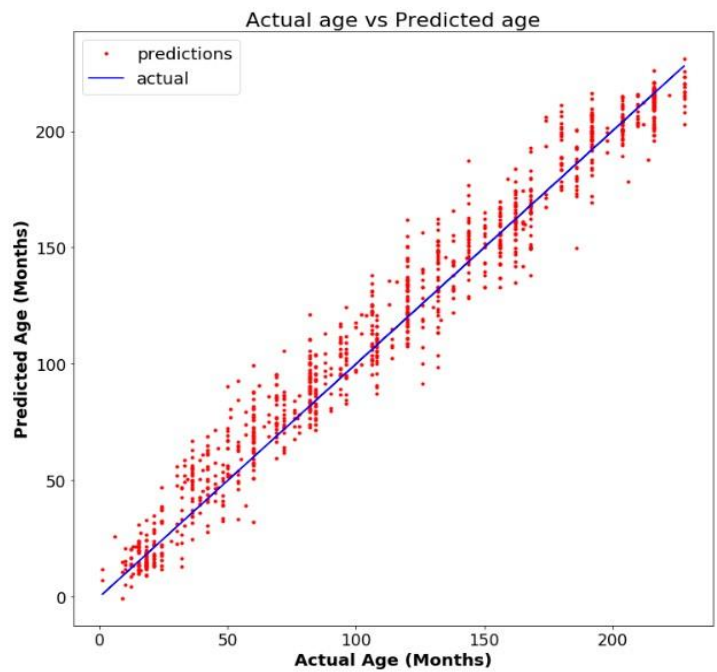

**Fig 26. Actual age vs. Predicted Age in months**

A mean average error (MAE) of 9.98 months is obtained on test data.

**Results of model training *using method 2*:**

Figure 27 shows the mean average error variation for the train and test images, and Figure 28 shows the model's prediction versus the actual data.

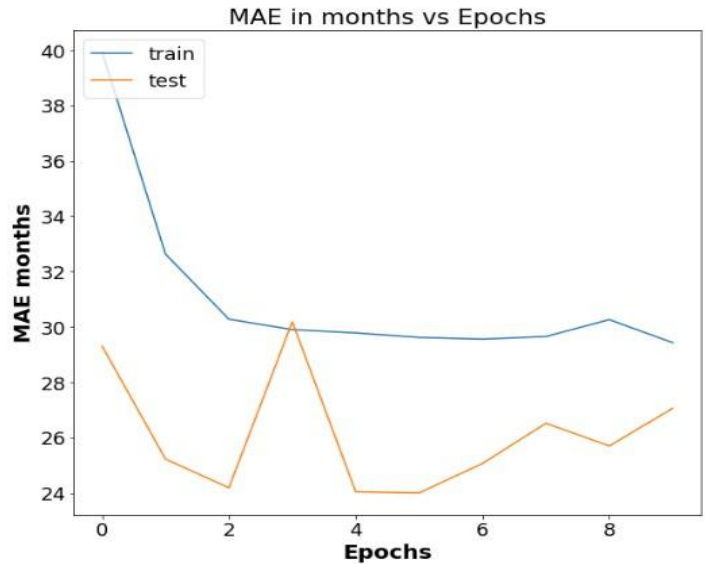

**Fig 27. MAE months vs. Epochs**

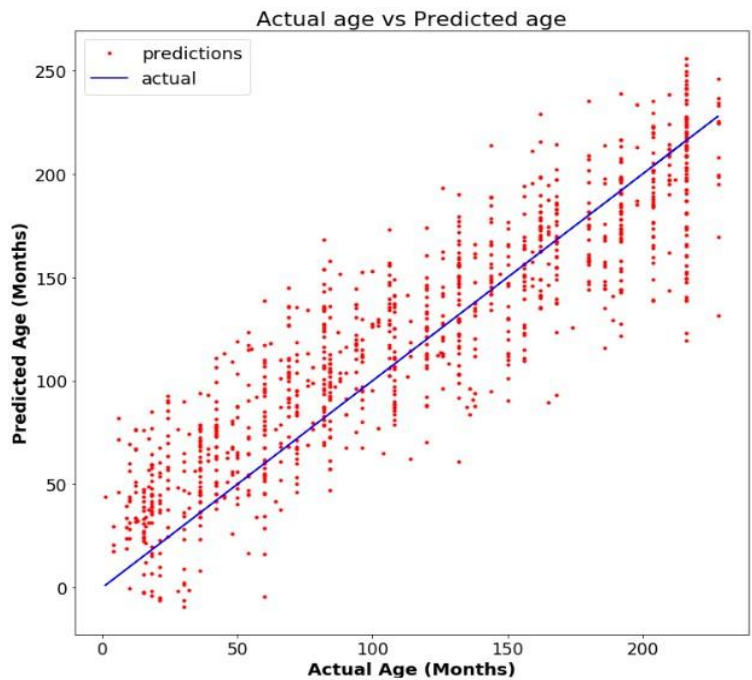

**Fig 28. Actual age vs. Predicted Age in months**

A mean average error (MAE) of 26.65 months is obtained on test data.

## 4. DISCUSSION

TABLE 1: Comparison of Mean average errors (MAE) of all the pre-trained models used

| Pretrained Model | MAE (in months) using method 1 | MAE (in months) using method 2 |
|---|---|---|
| VGG-16 | 14.00 | 29.51 |
| Inception V3 | 10.23 | 29.13 |
| MobileNet | 9.55 | 29.56 |
| XceptionNet | 9.98 | 26.65 |

According to Table 1, we can see that the MobileNet model using method 1, i.e., when all layers of the model were trained, performs the best with a mean average error (MAE) of 9.55 months on the test data. We can also see that XceptionNet performs better than the other three models when using method 2, i.e., only when the last few model layers were trained with an MAE of 26.65.

One of the significant limitations is the structure of the images obtained to train the models. Not all images are alike. Some of the hand X-ray images are twisted, inclined to either right or left; the background for some of the images is white, and some are black. Although various pre-processing techniques have been used, there are still some limitations considering all these scenarios. The number of X-ray images in the dataset of people aged between 140 months and 160 months is the highest, as shown in Figure 3. This leads to a bias in the model's prediction as it will have more data to train for the people of this age group and have higher accuracy in predicting the bone age of this people's age group than the others.

The prediction accuracy could be improved by training the model for more epochs, adding more examples to the training set, and ensuring that the dataset's population ratio and the x-ray images of people in different age groups remain uniform. This would help overcome the model bias towards a particular gender or age group.

## 5. CONCLUSION

An MAE of 9.55 was achieved using MobileNet employing method 1, the best among all 8 results achieved using the 4 different pre-trained models implemented using two methods. Future work can include improving the models' prediction using different ensemble techniques, i.e., combining the different pre-trained models to improve results.